# Equivalence analysis between Quasi-coarse-grained and Atomistic Simulations

Dong-Dong Jiang[1], Pei Wang[2], Jian-Li Shao[1,3*]

[1] National Key Laboratory of Explosive Science and Safety Protection, Beijing Institute of Technology, Beijing, 100081, China

[2] Institute of Applied Physics and Computational Mathematics, Beijing 100094, China

[3] Explosion Protection and Emergency Disposal Technology Engineering Research Center of the Ministry of Education, Beijing, 100039, China

## Abstract

In recent years, some mesoscale simulation methods based on the scaling of atomic potential functions, such as quasi-coarse-grained dynamics and coarse-grained dynamics, have shown promising results for modeling crystalline systems. However, this study presents evidence suggesting that the spatiotemporal trajectories of coarse-grained systems generated by such simulation methods exhibit a complete correspondence with those of specific molecular dynamics systems. In essence, current coarse-grained simulation methods involve a direct amplification of the results obtained from molecular dynamics simulations across spatial and temporal scales, yet they may lack the capability to adequately capture authentic scale effects. Consequently, the findings of related studies warrant careful re-evaluation. Furthermore, this study underscores the importance of not only verifying the consistency of mesoscale simulation methods with microscopic simulations but also meticulously assessing their capability to accurately forecast mesoscale physical phenomena.

*Corresponding author, email: * shao_jianli@bit.edu.cn

## I. Introduction

The importance of molecular dynamics (MD) simulations in advancing materials science and engineering cannot be overstated. These simulations offer a window into the fundamental nature of the world at the atomic scale, disregarding quantum effects. However, MD simulations have long grappled with two significant challenges. Firstly, their simulation outcomes hinge entirely upon the interatomic potential function, which has historically been scrutinized due to its reliance on model approximations and empirical parameters [1]. Secondly, MD simulations demand substantial computational resources, with the number of particles in macroscopic entities surpassing the computational capabilities of current supercomputers. The spatial and temporal scale of the largest all-atom MD simulations recorded to date are limited to the micrometer and microsecond [2], with the majority confined to the nanometer and nanosecond scales. This limitation makes it challenging to directly apply MD simulations to macroscopic dynamic problems across different length scales and to assess size effects between microscopic physical processes and macroscopic phenomena. In recent years, significant strides have been made in addressing the first challenge through the development of machine learning potential functions based on first principles. These advancements have enabled MD simulations with first-principles accuracy on scales exceeding a hundred million atoms, thereby significantly enhancing their reliability [3,4]. On the other hand, numerous mesoscopic simulation methods have emerged. Examples include coarse-grained MD for materials systems characterized by clear structural hierarchies, such as proteins and polymers [5–7], and dissipative particle dynamics for complex fluid systems [8,9]. These methods have been widely and successfully applied. However, for typical crystalline materials like metals and ceramics, the macroscopic mechanical behavior correlates with various microscopic-scale structural evolutions, such as dislocations and twinning [2,10,11]. In these microstructural evolution processes, interactions between individual atoms exert equally significant influences, distinguishing them from the selective bundling of particles in large molecular systems. Consequently, direct application of traditional coarse-graining methods to crystalline systems poses challenges.

Currently, mesoscopic simulation methods for crystalline systems can generally be categorized into two main types: mesoscopic theoretical simulation methods and concurrent multiscale coupling simulation methods. The former abstractly model actual material deformation and phase transition processes based on physical understanding, examples include phase-field dynamics [12–15] and discrete dislocation dynamics [16,17]. These methods heavily rely on empirical knowledge of physical laws and face challenges in handling various types of defects and interactions among different structures in practical processes. The latter category involves the use of full-atom molecular dynamics simulations to predict localized deformation and damage, coupled with macroscopic simulation methods like finite element analysis in other regions [18–20]. The exchange of physical information such as momentum and heat between the two regions enables their direct coupling, which has shown promising applications in problems like crack propagation [21,22]. However, when deformation and damage within the material are relatively uniform, such methods encounter significant difficulties. This is because the macroscopic simulation region struggles to accurately capture the thermodynamic effects of various physical processes like structural damage and phase transition. Particularly, numerous studies have indicated strong size effects and strain rate effects in processes such as damage [23,24], phase transition [25], and plastic deformation of crystalline materials [11,26], necessitating direct mesoscopic scale investigations into these processes.

To address the challenges encountered in simulating the mesoscale behavior of crystal materials, Dongare *et al*. [27] proposed a quasi-coarse-grained dynamic simulation method. This approach relies on representative particles to capture the overall dynamic behavior, leveraging the symmetry and periodicity inherent in crystal structures. By directly scaling the atomic potentials, the interactions between these representative particles are effectively modeled. This methodology offers the flexibility to scale the spatiotemporal scales of MD simulations by arbitrary factors, while still preserving crucial microscopic structural evolution details. It is considered a promising development route for multiscale simulations and has been applied to explore various mesoscale physical phenomena such as particle cold spraying [28], phase transitions [29], void collapse [30], and shock wave front evolution [31]. It has also found some applications in graphene and silicon [32]. Although the fundamental concept of this method originates from the structural characteristics of crystals, it is equally applicable to molten liquid systems. It has been successfully applied to the micro-spallation behavior in molten metals [33]. Furthermore, based on similar ideas to quasi-coarse-grained dynamics, there have emerged methods that directly aggregate the energy and mass of atoms within specific regions into single particles for coarse-grained simulations of crystals [34,35].

In summary, research on quasi-coarse-grained simulations of crystal materials represents a crucial direction in micro-meso-scale simulation. However, we will demonstrate here simply and reliably that the above quasi-coarse-grained dynamics are completely equivalent to full atomic-scale molecular dynamics simulations at the microscale. Therefore, quasi-coarse-grained simulations at any scale are equivalent to MD simulations.

## II. Theory and proof

### 2.1 The spatiotemporal correspondence between QCGD and MD simulations

To begin, let's review the basic principles of quasi-coarse-grained dynamics (QCGD). In both MD and QCGD simulation methods, a crucial step involves determining the potential energy $E_i$ of each particle (or atom) based on its local configuration, as delineated by Equations (1).

$$E_i = \Phi(\mathbf{r}_i) \tag{1}$$

The $\mathbf{r}_i$ denotes the set of relative positions between the particle i and all other particles, i.e., $\mathbf{r}_i = \{\mathbf{x}_j - \mathbf{x}_i \,|\, 1 \le j \le N, j \ne i\}$, $\mathbf{x}_i$ refers to the coordinate vector of particle i. N is the number of particles in the system. By calculating the partial derivative of the potential energy $E_i$ with respect to the coordinate $\mathbf{x}_i$, the force on the particle $\mathbf{F}_i$ can be obtained, i.e.:

$$\mathbf{F}_i = -\frac{\partial E_i}{\partial \mathbf{x}_i} \tag{2}$$

After obtaining the force acting on the partial, the position of partial i at the next moment can be obtained based on Newtonian mechanics theory and initial information on the partial position and velocity, gradually obtaining the spatiotemporal trajectories of the total system.

In quasi-coarse-grained dynamics, representative particles (R-particles) are proposed to characterize the average properties of all atoms within a specific region[27]. These R-particles possess identical neighborhood structural features as the initial atomic system. As depicted in Fig. 1, the all-atomic (AA) system with FCC structure in Fig. 1(a) can be represented by the quasi-coarse-grained (QCG) system composed of R-particles shown in Fig. 1(b). The QCG system has the same structure and orientation as the AA system, but with a larger lattice constant. The factor by which the lattice constant increases is termed the distance scaling parameter ($L_{cg}$). Consequently, each R-particle represents the average properties of $L_{cg}^3$ atoms. For convenience, we refer to a coarse-grained system with $L_{cg}$=N as an L'N' system. For example, when $L_{cg}$=2, as shown in Fig. 1b, it is called an L2 system.

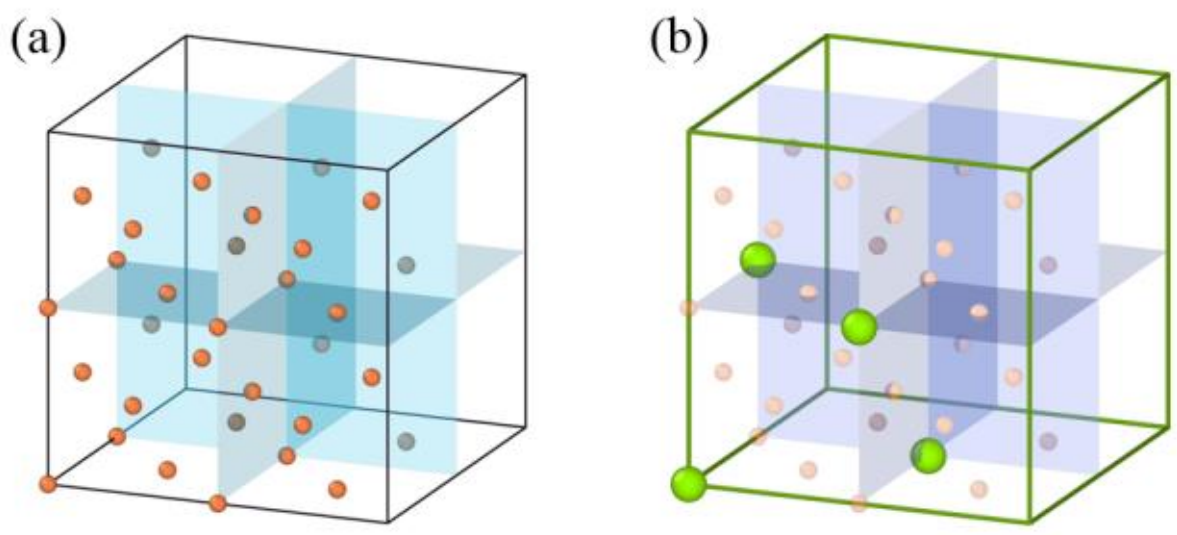

**Fig. 1** All-atomic system **(a)** and corresponding quasi-coarse-grained system **(b)**

After constructing the QCG system, another key aspect of QCGD simulations lies in defining the interaction potential between R-particles. Given the similarity in neighborhood structure between the QCG and AA systems, the potential function between R particles ($\Phi^R$) can be derived by directly scaling the interatomic potential function ($\Phi^{atom}$), as shown in Equation (3):

$$E_i^R = \Phi^R(\mathbf{r}_i^R) = \Phi^{atom}(\frac{\mathbf{r}_i^R}{A_{cg}}) \tag{3}$$

where $\mathbf{r}_i^R = \left\{ \mathbf{x}_j^R - \mathbf{x}_i^R \middle| 1 \le j \le N, j \ne i \right\}$, $\mathbf{x}_i^R$ refers to the coordinate vector of R-particles i. Notably, the function $\Phi^{atom}$ can assume various forms, encompassing potential functions such as pair potentials, many-body potentials, and even machine learning potentials. Based on this scaled potential function, R-particles at any strain can maintain the same energy characteristics as atoms in MD simulations. After introducing the total energy, stress, and temperature calculation methods related to the parameter $L_{cg}$ [27], the QCG system can predict thermodynamic properties such as temperature and pressure that are almost identical to those of the AA system. In addition, due to the complete preservation of the shape of the original interatomic potential function by the QCG potential function, it is completely consistent with MD simulations on various structural potential barriers, such as stacking fault energy and the energy of nucleation during phase transition. At the same time, in order to ensure the consistency of particle mass related dynamic characteristics such as wave velocity, thermal conductivity, and MD simulations, it is necessary to have the mass of the representative particle R equal to the atomic mass, i.e., $m^R = m^{atom}$, which is also consistent with the definition of the representative particle. The above QCGD simulation has been verified to be consistent with MD simulation in many properties such as melting point, melting rate, shock wave velocity, dislocation velocity, etc.

However, a new issue is that the capability of QCGD to describe scale effects as the distance scaling parameter increases remains unproven. Here we consider an arbitrary quasi-coarse-grained system, where the set of coordinates of all R atoms at time t is denoted as $\mathbf{R}_t^R = \left\{ \mathbf{x}_i^R \middle| 1 \le i \le N \right\}$. As shown in Fig. 2, there exists a corresponding all-atom system, a direct scaled down of the CG system in space, where the set of coordinates of all atoms is denoted as $\mathbf{R}_t^{atom}$, i.e.

$$\mathbf{R}_t^{atom} = \frac{\mathbf{R}_t^R}{L_{cg}} \tag{4}$$

Therefore, for R-particle i and its corresponding atom i, there are:

$$\mathbf{r}_i^{atom} = \frac{\mathbf{r}_i^R}{L_{cg}} \tag{5}$$

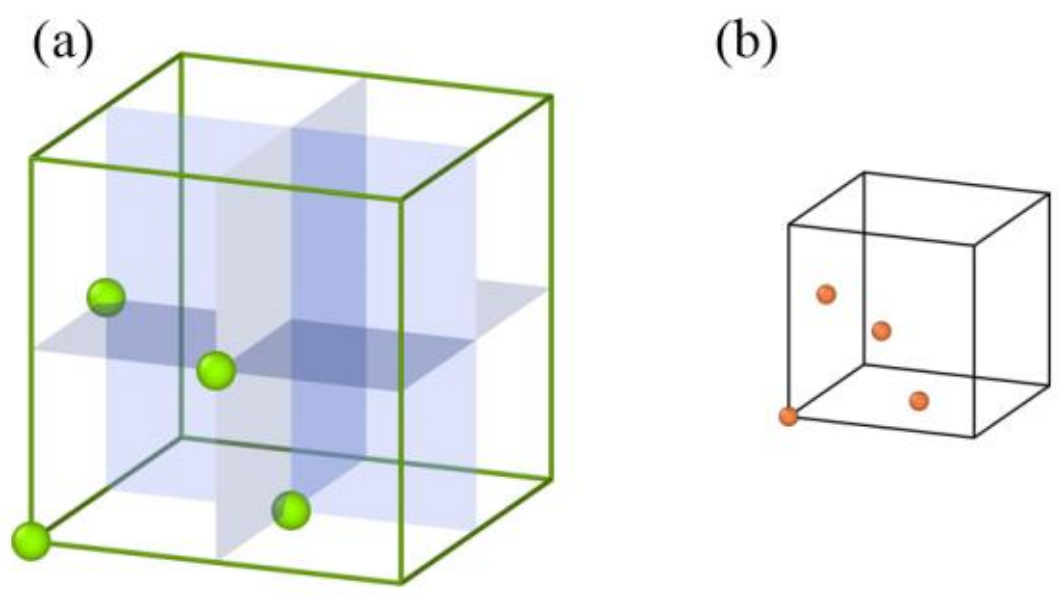

**Fig. 2** Quasi-coarse-grained system (a) and corresponding scaled down all-atom system (b)

And the atom velocities of this atomic system are same to the R-particle velocities:

$$\mathbf{V}_t^{atom} = \mathbf{V}_t^R \tag{6}$$

where $\mathbf{V}_t = \left\{ \mathbf{v}_i \middle| 1 \le i \le N \right\}$.Under the above initial conditions, consider the independent subsequent evolution of these two systems. Combining equations (1)-(5), it can be concluded that for any R-particle i and its corresponding atom i, there are:

$$\mathbf{F}_i^R = \frac{\mathbf{F}_i^{atom}}{L_{cg}} \tag{7}$$

Combining $m^R = m^{atom}$ , there is $\mathbf{A}_t^R = \frac{\mathbf{A}_t^{atom}}{L_{cg}}$ , where $\mathbf{A}_t = \left\{ \mathbf{a}_i^t \middle| 1 \le i \le N \right\}$ . After Δt time, the set of coordinates of all atoms is:

$$\mathbf{R}_{t+\Delta t}^{atom} = \mathbf{R}_t^{atom} + \mathbf{V}_t^{atom}\Delta t + \frac{1}{2}\mathbf{A}_t^{atom}\Delta t^2 \tag{8}$$

Correspondingly, after $L_{cg}\Delta t$ time, the quasi coarse-grained system has:

$$\mathbf{R}_{t+L_{cg}\Delta t}^{R} = \mathbf{R}_t^{R} + L_{cg}\mathbf{V}_t^{R}\Delta t + \frac{1}{2}L_{cg}^2\mathbf{A}_t^{R}\Delta t^2 \tag{9}$$

Substituting equations (4) and (6) yields:

$$\mathbf{R}_{t+L_{cg}\Delta t}^{R} = L_{cg}\mathbf{R}_{t+\Delta t}^{atom} \tag{10}$$

At the same time, it is easy to obtain:

$$\mathbf{V}_{t+L_{cg}\Delta t}^{R} = \mathbf{V}_{t+\Delta t}^{atom} \tag{11}$$

This implies that for any QCG system, there exists a corresponding AA system, and the two systems show complete correspondence in their subsequent spatiotemporal evolution. Hence, it stands to reason that a QCGD simulation can reproduce all physical properties and evolutionary behaviors observed in molecular dynamics simulations under identical stress and temperature conditions. Nevertheless, this also rules out the possibility of utilizing increase coarsening level to capture any mesoscale behavior.

### 2.2 The spatiotemporal correspondence between CGMD and MD simulations

In CGMD simulations, unlike QCGD simulations, each coarse-grained particle no longer represents the average property of atoms within a specific region, but directly replaces all atoms in the region. Therefore, its particle mass is the sum of atomic masses within the region, i.e., $m^{CG} = L_{cg}^3 m^{atom}$ . The relationship between the potential energy of coarse-grained particles and the potential energy of atoms is:

$$E_i^{CG} = L_{cg}^3\Phi^{atom}(\frac{\mathbf{r}_i}{L_{cg}}) \tag{12}$$

The CGMD simulation method, in contrast to the QCGD approach, naturally preserves the same mass density and total energy density as the atomic system without necessitating additional conversions. However, akin to QCGD, for any coarse-grained system, there exists a corresponding all-atom system, conforming to the relationships articulated in Equation 4 and Equation 5. Specifically, for any $i$ particle and its corresponding $i$ atom, $\mathbf{r}_i^{CG} = L_{cg}\mathbf{r}_i^{atom}$ . Combining Equations 1, 2, and 12, we obtain $\mathbf{F}_i^{CG} = L_{cg}^2\mathbf{F}_i^{atom}$ . Taking into account the correspondence between the mass of coarse-grained particles and that of atoms, similar to QCMD, we still find $\mathbf{A}^R = \frac{\mathbf{A}^{atom}}{L_{cg}}$. Thus, akin to QCGD, subsequent particle trajectories in CGMD simulations align precisely with the atomic trajectories of the all-atom system. The related thermodynamic properties and microscopic physical attributes are also entirely consistent. Evidently, CGMD lacks fundamental disparity from QCGD.

## III. Example verification

### 3.1 Uniaxial compression of polycrystalline copper

In practical simulations, the time step of the QCG system can be set to $L_{cg}$ times that of the MD simulation system. By combining this with Equation 9, it is theoretically observed that the particle positions and velocities of the QCG system align perfectly with those of the AA system after the same number of steps. It is important to ensure that the corrections to particle positions and velocities are consistently applied at each step. Under this condition, artificial loading such as impacts, tensile/compressive deformations, and ensemble corrections will not disrupt this alignment. Taking the example of uniaxial compression of polycrystalline copper with constant volume, the time step for the AA system is 2 fs, and the strain rate is $1\times10^{10}$ /s. When $L_{cg}$ is 10, the time step becomes 20 fs, corresponding to a compression strain rate of $1\times10^{9}$ /s. When $L_{cg}$ increase to 100, the strain rate further reduces to $1\times10^{8}$ /s. This setting ensures that the three models are compressed by the same proportion at each time step. The specific simulation results are depicted in Figure 3. In the initial thousands of steps, the three simulations exhibit complete correspondence, with stress-strain curves perfectly overlapping. Even the positions, velocities, and local structures of each atom are identical (as shown in the left images of Figure 3b and c), consistent with our previous demonstration. However, as the simulation progresses, at strain values greater than 0.1, the stress-strain curves of the three models no longer overlap completely. Nevertheless, they still follow consistent evolution patterns. From the comparison of particle neighborhood structures in the right images of Figure 3b and c, it can be observed that the particle structure characteristics of the AA system and the L100 QCG system remain consistent, albeit specific dislocation and stacking fault positions show some differences. This discrepancy arises from floating-point math not being associative and the chaotic nature of MD simulations. During the QCGD simulation process, scaling the potential function and system coordinates can introduce minimal numerical errors, causing slight differences and eventual divergence of molecular dynamics phase space trajectories within a few thousand timesteps. However, the statistical properties of the two runs (e.g., average energy or temperature) should remain consistent. Therefore, these minor differences resulting from the chaotic nature of the system do not affect the correspondence between the two systems as demonstrated above. In summary, the simulations above demonstrate that although the strain rates of these three systems differ by a factor of 100, the stress-strain curves are nearly identical, fully proving the consistency between all-atom simulations and QCGD simulations after temporal and spatial scaling.

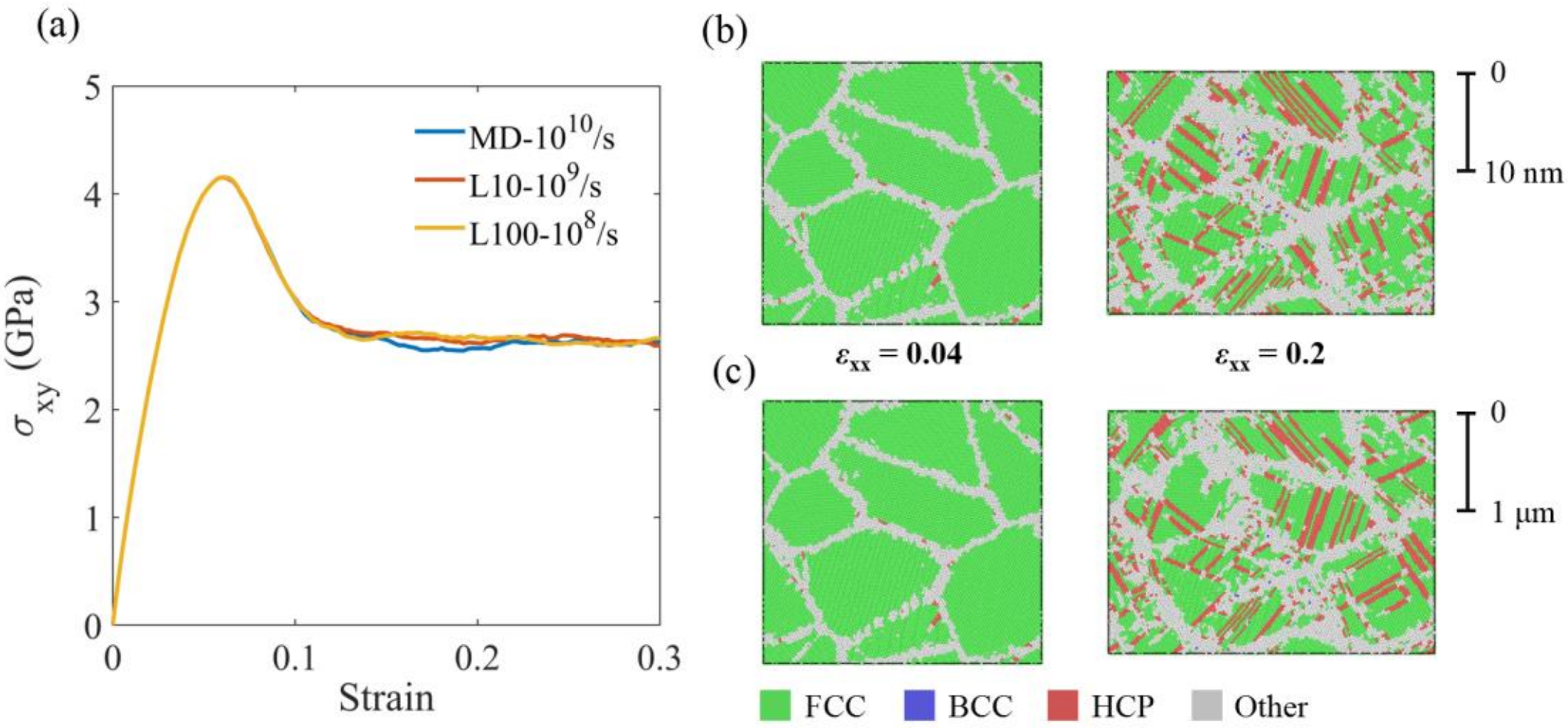

**Fig. 3 (a)** Evolution of shear stress during compression process; **(b)** Atomic local configurations obtained from molecular dynamics simulations at different strains; **(c)** Particle local configurations obtained from coarse-grained dynamics simulations at different strains.

### 3.2 Shock loading of polycrystalline aluminum

Below, we illustrate how the QCGD and CGMD simulation methods have demonstrated misleading scale effects, using the example of the shock loading process of polycrystalline aluminum. For instance, when investigating the impact of grain size on shock wave profiles, we vary the coarsening level to generate models of varying grain sizes, similar to a previous study using QCGD method [31]. The configurations of all simulation groups are outlined in Table 1, and the resulting simulations are depicted in Fig. 4. Initially, MD simulations and QCGD simulations with different coarsening level were employed to calculate the shock loading process of the same model. Fig. 4(a) and (b) show the characteristics of shock waveforms and the attenuation of precursor amplitudes with increasing shock depth. It can be observed that the results of QCGD simulations are not significantly different from those of MD simulations. This similarity is often used to demonstrate that quasi-coarse-grained methods are reliable approximations of MD simulations.

**Table 1.** Initial model setup for polycrystalline aluminum shock simulation.

| Num | Coarsening Level | Particle Count | Model size (nm$^3$) | Grain Sizes (nm) |
|---|---|---|---|---|
| 1 | L1 | 86259771 | 60×60×400 | 25.5 |
| 2 | L2 | 10770843 | 60×60×400 | 25.5 |
| 3 | L4 | 1345193 | 60×60×400 | 25.5 |
| 4 | L1 | 10770843 | 30×30×200 | 12.8 |
| 5 | L2 | 10770843 | 60×60×400 | 25.5 |
| 6 | L4 | 10770843 | 120×120×800 | 51.0 |
| 7 | L8 | 10770843 | 240×240×1600 | 102 |
| 8 | L16 | 10770843 | 480×480×3200 | 204 |

Furthermore, by adjusting the coarsening level, models with the same number of particles can be used to study the influence of grain size on the attenuation behavior of elastic precursor waves. As shown in Fig. 4(c), the results indicate that larger grain sizes lead to slower attenuation of elastic precursor waves. However, as demonstrated earlier in this paper, the simulation results of different grain sizes are actually just different manifestations of simulations of the same all-atomic model, rather than differences in the physical properties of systems brought about by different grain sizes. If the shock depth on the abscissa in Fig. 4(c) is changed to the proportion of the shock depth to the entire model, as shown in Fig. 4(d), it can be observed that these groups of simulations exhibit completely consistent patterns, with only minor differences due to random thermal fluctuations in the system.

The above results further confirm the complete correspondence between the QCGD simulations results and the corresponding all-atom simulation results. However, due to the direct amplification of MD simulation results in both time and space dimensions, certain physical phenomena may exhibit false size-effect characteristics, potentially leading researchers to overlook this crucial issue.

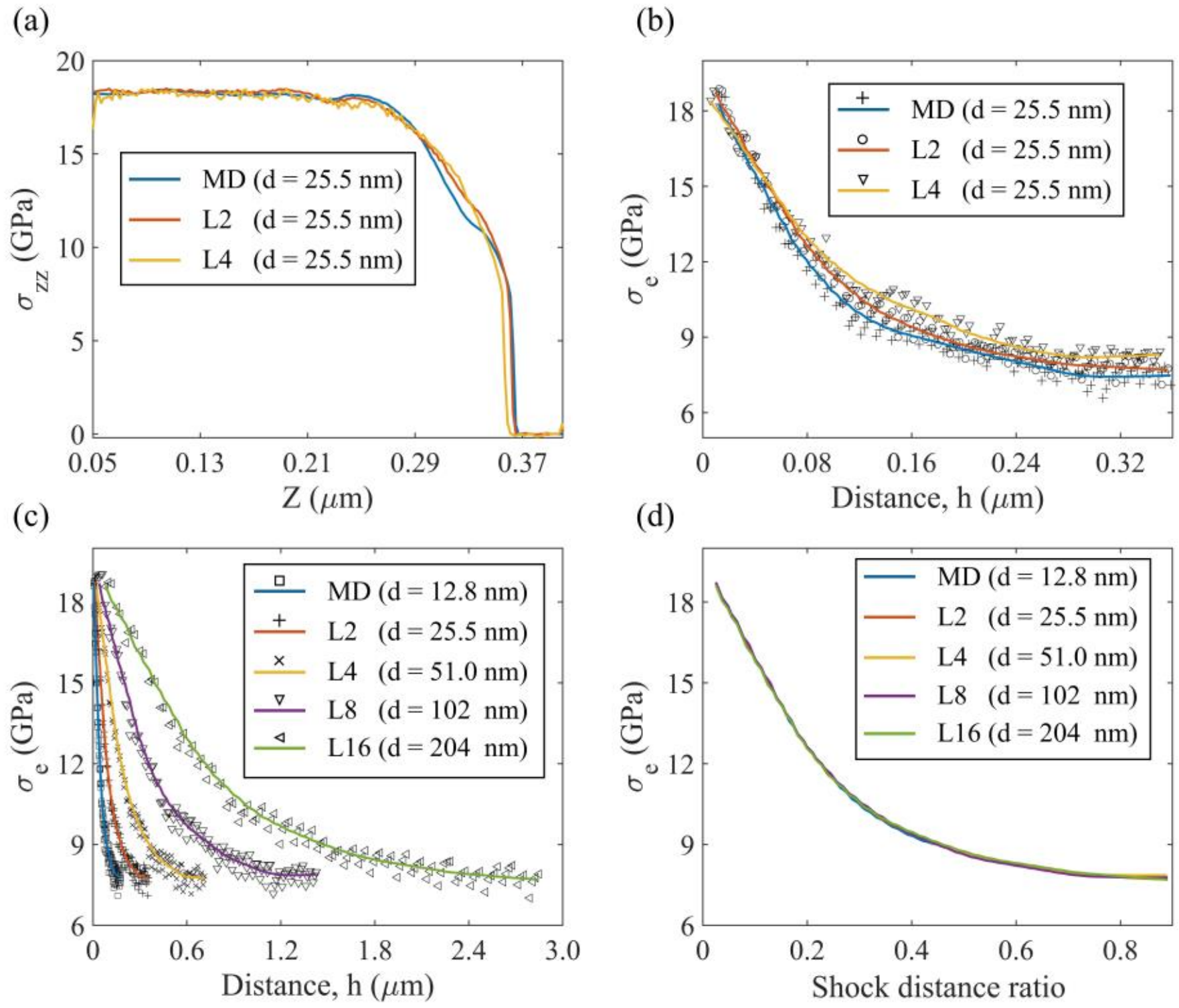

Fig. 4 (a) Shock wave profiles under different coarsening level; (b) Attenuation of elastic precursor amplitude with shock depth under varying coarsening level; (c) Attenuation of elastic precursor amplitude with shock depth under different grain sizes; (d) Attenuation of elastic precursor amplitude with shock ratio (ratio of shock depth to sample thickness) under different grain sizes.

## IV. Conclusions

In this study, we demonstrate that there is no fundamental difference between the computational results of quasi-coarse-grained dynamics (QCGD) based on atomic potential scaling and those of molecular dynamics with the same number of particles. The strategy of adjusting the degree of coarse-graining to study mesoscopic effects is therefore not feasible. As a result, although the QCGD method has been widely applied [28–33] , we must recognize that the research results based on the QCGD method do not actually have real physical significance. For crystal systems such as metal materials, there is currently no mesoscale simulation method that can naturally describe the microstructural evolution behavior.

Indeed, virtually all multiscale simulation approaches stemming from microscopic simulations encounter this challenge [36–38]. In addition to verifying the consistency between mesoscopic simulation results and their full atomic counterparts, it is imperative to meticulously evaluate their capacity to accurately capture mesoscale physical phenomena. Considering that the reduction of system degrees of freedom during coarse-graining inevitably leads to the loss of physical information, a potential solution lies in compensating for this loss by incorporating intrinsic physical quantities onto coarse-grained particles. Additionally, it may be necessary to introduce corrective terms, such as dissipation and random forces, into the dynamics of the coarse-grained model.

**Funding**: This work is supported by the National Key R&D Program of China (Grant No. 2021YFB3802300).